\documentclass[aip,reprint,amsfonts,amsmath,amssymb,floatfix,citeautoscript,groupedaddress]{revtex4-1}
\usepackage{graphicx}
\usepackage{dcolumn}
\usepackage{bm}
\usepackage{parskip}
\usepackage{enumerate}
\usepackage{color}
\usepackage{psfrag}
\usepackage{upgreek}
\setcitestyle{numbers,square}
\setcitestyle{notesep={ }}

\usepackage{lineno}

\begin{document}

\title{
phase behavior of hard circular arcs
}

\author{Juan Pedro Ram\'irez Gonz\'alez}
\affiliation{
Departamento de F\'isica Te\'orica de la Materia Condensada,  \\
Universidad Aut\'onoma de Madrid,
Ciudad Universitaria de Cantoblanco, \\
E-28049 Madrid, Spain
}

\author{Giorgio Cinacchi}
\affiliation{
Departamento de F\'isica Te\'orica de la Materia Condensada,  \\
Instituto de F\'isica  de la Materia Condensada (IFIMAC),    \\
Instituto de Ciencias de Materiales ``Nicol\'{a}s Cabrera'', \\
Universidad Aut\'onoma de Madrid,
Ciudad Universitaria de Cantoblanco, \\
E-28049 Madrid, Spain
}

\date{\today}

\begin{abstract}
By using Monte Carlo numerical simulation,
this work investigates the phase behavior of
systems of hard infinitesimally--thin circular arcs,
from an aperture angle $\uptheta \rightarrow 0$
to an aperture angle $\uptheta \rightarrow 2 \uppi$,
in the two--dimensional Euclidean space.
Except 
in the isotropic phase at lower density and
in the (quasi)nematic phase, 
in the other phases that form, 
including the isotropic phase at higher density, 
hard infinitesimally--thin circular arcs auto--assemble to form clusters.
These clusters are  
either filamentous, for smaller values of $\uptheta$, 
or roundish, for larger values of $\uptheta$.
Provided density is sufficiently high,
the filaments lengthen, merge and straighten 
to finally produce 
a filamentary phase
while 
the roundels compact and dispose themselves with 
their centres of mass at the sites of a triangular lattice 
to finally produce 
a cluster hexagonal phase.
\end{abstract}

\maketitle

\section{introduction and motivation}
\label{introduction}

Many aspects of the physics of 
[(soft--)condensed] states of matter \cite{chaikin}
can be fruitfully investigated by
resorting to 
basic simple systems 
of hard particles 
\cite{torquato}.
Such particles interact between them solely via
infinitely repulsive short--range interactions
preventing them from intersecting.
Thus, entropy is,
on varying number density $\uprho$,
the sole physical magnitude that determines 
the 
phase behavior of such systems.
Yet, 
the infinitely repulsive short--range interactions provenly suffice for 
causing 
multiple fluid and solid states of matter to occur 
in systems 
of particles interacting via them.
This fact together with 
their omnipresence across 
length scales
justify the interest in systems 
of hard particles. 

The hard sphere 
is basic to
the broad condensed matter and statistical physics.
Systems of hard spheres have been extensively investigated 
with different composition and under a variety of conditions: 
A vast bibliography has been accumulated \cite{torquato}. 

In the course of the last fifty years,
the investigation has been progressively expanded to
systems of hard non-spherical particles \cite{torquato}.
They form
more complex instances of 
the fluid and solid states of matter 
that systems of hard spheres already exhibit \cite{chaikin,torquato}
along with genuinely new 
plastic--crystalline \cite{torquato,crisplastici} 
and liquid--crystalline \cite{chaikin,torquato,crisliquidi,mvmrrev,allenrev} 
states of matter.
The investigation on 
this progressively expanding variety of 
systems of hard non--spherical particles 
has actually shown 
how finely the hard-particle shape 
may determine the system phase behavior \cite{torquato}. 

The majority of these hard non--spherical particles are convex 
\cite{torquato}.
If non--sphericity causes 
genuinely new states of matter to occur,
non--convexity might promote special instances of 
fluid and solid states of matter. 
These states of matter might be difficultly achievable
or entirely precluded
in systems of hard, 
convex however dexterously shaped, 
particles.

\begin{figure}
\centering
\includegraphics[scale=2]{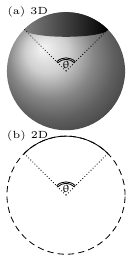}
\caption{(a) In three dimensions (3D), 
given a  spherical surface (shaded lighter gray), 
a spherical cap(sid) (shaded darker gray) is 
a portion of it whose any arc subtends an angle $\uptheta$.
(b) In two dimensions (2D), 
given a circumference (discontinuous line), 
an arc (continuous line) is 
a portion of it that subtends an angle $\uptheta$.} 
\label{lafigura0}
\end{figure}

Out of the minority of 
hard concave particles that 
have been considered thus far \cite{escobedo},
one is the hard spherical cap(sid) \cite{calotta}.
It consists of 
that portion of 
a spherical surface 
in the three--dimensional Euclidean space $\mathbb{R}^3$ 
whose any arc
subtends an angle $\uptheta \in [0,2 \uppi]$ [Fig. \ref{lafigura0} (a)]. 
These hard infinitesimally--thin curved  particles interpolate
between the hard infinitesimally--thin disc,
corresponding to $\uptheta = 0$, and the hard sphere,
corresponding to $\uptheta =2 \uppi$. 
In the latter decade, 
systems of hard spherical caps with $\uptheta \in [0,\uppi]$
were investigated \cite{calotta}.
Their phase behavior features
purely entropy--driven cluster columnar and cluster isotropic phases. 
Since similar, ``contact-lens--like'', colloidal particles have been synthesised \cite{pine}, 
these theoretical predictions 
could be experimentally tested.

\begin{figure}
\includegraphics{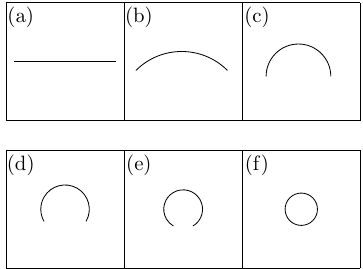}
\caption{Examples of circular arcs. They have the same length and different subtended angle $\uptheta$.
Those in the top row are minor:
(a) $\uptheta = 0$; 
(b) $\uptheta =  \displaystyle \frac{\uppi}{2}$; 
(c) $\uptheta=\uppi$.
Those in the bottom row are major:
(d) $\uptheta = \displaystyle \frac{4}{3} \uppi$; 
(e) $\uptheta = \displaystyle \frac{5}{3} \uppi$; 
(f) $\uptheta=2 \uppi$.
}
\label{lafigura1}
\end{figure}

Before complementing the investigation on 
systems of hard spherical caps \cite{calotta} 
by investigating systems of hard spherical capsids with $\uptheta \in (\uppi,2 \uppi]$,
it has seemed opportune to dedicate the present investigation to
the two--dimensionally analogous problem:
The complete phase behavior of 
systems of 
hard infinitesimally--thin circular arcs
in the two--dimensional Euclidean space $\mathbb{R}^2$ 
that subtend an angle $\uptheta \in [0, 2 \uppi]$
[Fig. \ref{lafigura0} (b)].
This class of hard curved particles interpolates between 
the hard segment, corresponding to $\uptheta = 0$, and
the hard circle, corresponding to $\uptheta = 2 \uppi$;
it can be divided into 
the sub--class of hard infinitesimally--thin minor circular arcs, 
from $\uptheta = 0$ up to $\uptheta=\uppi$, and 
the sub--class of hard infinitesimally--thin major circular arcs,
from $\uptheta = \uppi^{+}$ up to $\uptheta=2 \uppi$ (Fig. \ref{lafigura1}).

In addition to the utility of 
addressing the same type of physical problem across different dimensions,
there is another
motivation to investigate 
the complete phase behavior of 
systems of hard infinitesimally--thin circular arcs. 
It is the desire of exploring
whether the recently constructed densest--known packings of 
hard infinitesimally--thin major circular arcs \cite{hcadkp} 
or
sub--optimal versions of them 
can spontaneously form.
These densest--known packings 
consist of compact circular 
clusters that comprise        
\begin{equation}
\displaystyle 
\mathtt{n} = \left [ \frac{2 \uppi}{\uptheta -\uppi} \right ] 
\label{equxn}
\end{equation}
\cite{spiega} 
(anti)clockwise intertwining 
hard infinitesimally--thin major circular arcs
and dispose themselves with their centres of mass at 
the sites of 
a triangular lattice \cite{hcadkp}.
It should be probed whether 
a similar cluster 
phase 
will finally emerge out of 
a competition with 
the other phases that 
systems of hard infinitesimally--thin circular arcs
form.

To characterise these phases, 
a set of order parameters and correlation functions was considered (Section \ref{opcf}). 
These structural descriptors were calculated by 
statistically analysing 
the configurations that were saved and stored 
in the course of 
isobaric(--isothermal) Monte Carlo numerical simulations 
\cite{MCorigin,MCWood,MClibri} 
(Section \ref{mc}).
Out of the phases that the resulting phase diagram features, 
one is 
that cluster 
phase.
Provided $\uprho$ is sufficiently high, 
it forms in systems of 
hard infinitesimally--thin (quasi) major circular arcs. 
This phase constitutes the spontaneous, 
though sub--optimal, 
version of the densest--known packings that 
have been recently determined \cite{hcadkp} 
(Section \ref{phasediagram}).
While sketching this phase diagram,
a few traits of the phases that it features and of 
the transitions between them emerge. 
that 
would require 
as many dedicated theoretical investigations.
It is hoped that the present results 
stimulate these theoretical investigations along with 
the preparation of colloidal or granular thin-circular-arc--shaped particles and 
the ensuing experimental investigation of systems of them 
(Section \ref{conclusion}).

\section{methods}

\subsection{order parameters and correlation functions}
\label{opcf}


Certain of the 
order parameters and correlation functions are 
ordinary and prefigurable 
based on 
the non-sphericity and
(generally $\mathsf{D}_{{\mathrm {1}}}$ \cite{O2}) symmetry 
of the present hard particles 
and 
the abundant previous work on 
systems of hard (non-)spherical particles \cite{torquato}.

\begin{figure}
\centering
\includegraphics{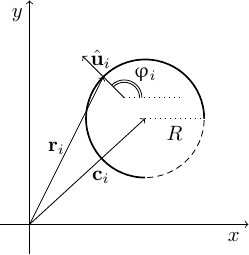}
\caption{Example of a circular arc $i$ (continuous line),
in a Cartesian $(x,y)$ reference frame with 
several quantities that define its mechanical state and 
enter the definition of order parameters and correlation functions:
$\mathbf{r}_i$, the vector of the position of its vertex;
${\hat{\mathbf{u}}}_i$, the unit vector of its orientation,
which lies on the direction joining the centre of
its parent circle with the vertex 
and
forms an angle $\upvarphi_i$ with the $x$ axis;
$\mathbf{c}_i$, the vector of the position of the centre
of its parent circle (discontinuous line)  whose radius is $R$. 
}
\label{lafigura2}
\end{figure}

The most basic correlation function is
the positional pair--correlation function which,
in an uniform, or treated as if it were such, system of 
$N$ particles is usually indicated as $g(r)$. 
It can be defined as:
\begin{equation}
g(r) = \frac{1}{N} 
\left \langle \frac{1}{\uprho}
\sum_{i=1}^N \sum_{j \ne i}^N \updelta 
\left ( \left | {\mathbf{r}}_j - {\mathbf{r}}_i \right | - r \right )
\right \rangle 
\label{gdr}
\end{equation}
with $\left \langle \, \right \rangle$ 
signifying a mean over configurations,
$\updelta()$ the usual $\updelta$-function and
$\mathbf{r}_i$ the position of 
the centroid of particle $i$;
presently, this centroid coincides with 
the vertex  of the circular arc $i$  
(Fig. \ref{lafigura2}). 

One order parameter that  the symmetry of
the present hard particles simply suggests is
the \textsl{polar} order parameter \texttt{S}$_1$.
It can be defined as:
\begin{equation}
\mathtt{S}_1 = 
\frac{1}{N} 
\left \langle 
\left | \sum_{i=1}^N \mathsf{e}^{\mathsf{i}\upvarphi_i} \right|
\right \rangle = 
\frac{1}{N} 
\left \langle 
\left | \sum_{i=1}^N {\hat{\mathbf{u}}}_i \right| \right \rangle \,
\end{equation} 
with 
${\hat{\mathbf{u}}}_i = (\cos \upvarphi_i \,,\, \sin \upvarphi_i)$ 
the unit vector  
along the symmetry axis of the circular arc $i$  
(Fig. \ref{lafigura2}).

The non-sphericity of 
the present hard particles 
suggests 
the calculation of 
the \textsl{nematic} order parameter $\mathtt{S}_2$.
It can be defined as:
\begin{equation}
\mathtt{S}_2 = \frac{1}{N} \left \langle 
\left | \sum_{i=1}^N \mathsf{e}^{\mathsf{i} 2 \upvarphi_i } \right | \right \rangle
= \left \langle 
\frac{1}{N} \sum_{i=1}^N
\left[
2 
\left( 
{\hat{\mathbf{u}}}_i\cdot{\hat{\mathit{n}}} 
\right )^2  - 1 \right ] 
\right \rangle \,
\end{equation} 
with ${\hat{\mathit{n}}}$ the nematic director, i.e., 
the direction along which 
the orientation of a circular arc more probably aligns
\cite{vieillard}.

The two order parameters 
$\texttt{S}_1$ and $\texttt{S}_2$ 
would serve to establish whether and of which type
a phase possesses orientational order.
In actuality, associated to each of these order parameters is there
an orientational pair--correlation function that 
provides significantly more information.
The two respective correlation functions, 
${\mathcal G}_1(r)$ and ${\mathcal G}_2(r)$, 
are defined as:
\begin{widetext}
\begin{equation}
{\mathcal G}_1(r) = \left \langle 
\frac{\sum_{i=1}^N \sum_{j \ne i}^N 
\updelta 
\left ( \left | {\mathbf{r}}_j - {\mathbf{r}}_i \right | - r \right )
\left ( 
{\hat{\mathbf{u}}}_i 
\left ( {\mathbf{r}}_i \right ) 
\cdot 
{\hat{\mathbf{u}}}_j 
\left ( {\mathbf{r}}_j \right )
\right )}
{\sum_{i=1}^N \sum_{j \ne i}^N 
\updelta 
\left ( \left | {\mathbf{r}}_j - {\mathbf{r}}_i \right | - r \right )}   \right \rangle \,;
\label{laG1orienta}
\end{equation}

\begin{equation}
{\mathcal G}_2(r) = \left \langle 
\frac{\sum_{i=1}^N \sum_{j \ne i}^N 
\updelta 
\left ( \left | {\mathbf{r}}_j - {\mathbf{r}}_i \right | - r \right )
\left( 2 \left (  
{\hat{\mathbf{u}}}_i \left ( {\mathbf{r}}_i \right ) 
\cdot 
{\hat{\mathbf{u}}}_j \left ( {\mathbf{r}}_j \right )
\right )^2 - 1  \right)
} 
{\sum_{i=1}^N \sum_{j \ne i}^N 
\updelta 
\left ( \left | {\mathbf{r}}_j - {\mathbf{r}}_i \right | - r \right )}   
\right \rangle \,.
\label{laG2orienta}
\end{equation}
\end{widetext}
Not only would the values of $\mathtt{S}_1$ and
$\mathtt{S}_2$ be obtainable from 
the $r \rightarrow \infty$
limit of, respectively, 
${\mathcal G}_1(r)$ and ${\mathcal G}_2(r)$
but also the calculation of
orientational pair--correlation functions allows one 
to more profoundly characterise the orientational order
of a phase. 
In fact, 
the
possible tendency of two particles to mutually align can be 
characterised for
any distance separating them and
the way by which that long-distance limit is approached
can be probed
\cite{comment1}.

The possible formation of anisotropic phases suggests  
the definition of additional orientational pair--correlation functions whose argument is 
the inter-particle distance vector that is resolved along a certain specific direction. 
In particular, one can consider the orientational pair--correlation function 
${\mathcal G}_{1,\perp}^{\hat{\mathbf{u}}}(r_{\perp})$ defined as: 
\begin{widetext}
\begin{equation}
{\mathcal G}_{1,\perp}^{\hat{\mathbf{u}}} (r_{\perp}) = 
\left \langle 
\frac{\sum_{i=1}^N \sum_{j \ne i}^N 
\updelta 
\left ( 
\left| {\mathbf{r}}_j - {\mathbf{r}}_i - 
\left ( {\mathbf{r}}_j - {\mathbf{r}}_i \right ) \cdot 
{\hat{\mathbf{u}}}_i {\hat{\mathbf{u}}}_i \right|  - 
r_{\perp} \right )
\left ( 
{\hat{\mathbf{u}}}_i 
\left ( {\mathbf{r}}_i \right ) 
\cdot 
{\hat{\mathbf{u}}}_j 
\left ( {\mathbf{r}}_j \right )
\right )}
{\sum_{i=1}^N \sum_{j \ne i}^N 
\updelta 
\left ( 
\left| {\mathbf{r}}_j - {\mathbf{r}}_i - 
\left ( {\mathbf{r}}_j - {\mathbf{r}}_i \right ) \cdot 
{\hat{\mathbf{u}}}_i {\hat{\mathbf{u}}}_i \right|  - 
r_{\perp} 
\right ) }
\right \rangle \,.
\label{laG1orientaparallela}
\end{equation}
\end{widetext}
It probes 
the polar orientational correlations between two particles separated by 
a distance vector that is resolved along 
the direction perpendicular to the orientation of one of them.

The particular nature of the present hard particles
and the fact that systems of them may form,
in addition to the isotropic and (quasi)nematic \cite{quasi} phases, 
distinctive phases suggest special order parameters and correlation functions. 

The particular nature of the present hard particles
suggests to probe        
the positional correlation between 
a pair of them
in terms of 
the centres of their parent circles.
The definition of 
the corresponding  pair--correlation function $G(c)$ 
parallels that of $g(r)$ in Eq. \ref{gdr}: 
\begin{equation}
G(c) = \frac{1}{N} 
\left \langle \frac{1}{\uprho}
\sum_{i=1}^N \sum_{j \ne i}^N \updelta 
\left ( \left | {\mathbf{c}}_j - {\mathbf{c}}_i \right | - c \right )
\right \rangle 
\label{Gdc}
\end{equation}
with ${\mathbf{c}_i}$ the position of
the centre of the parent circle of particle $i$
(Fig. \ref{lafigura2})
\cite{comment2}.

\begin{figure}
\centering
\includegraphics[scale=0.7]{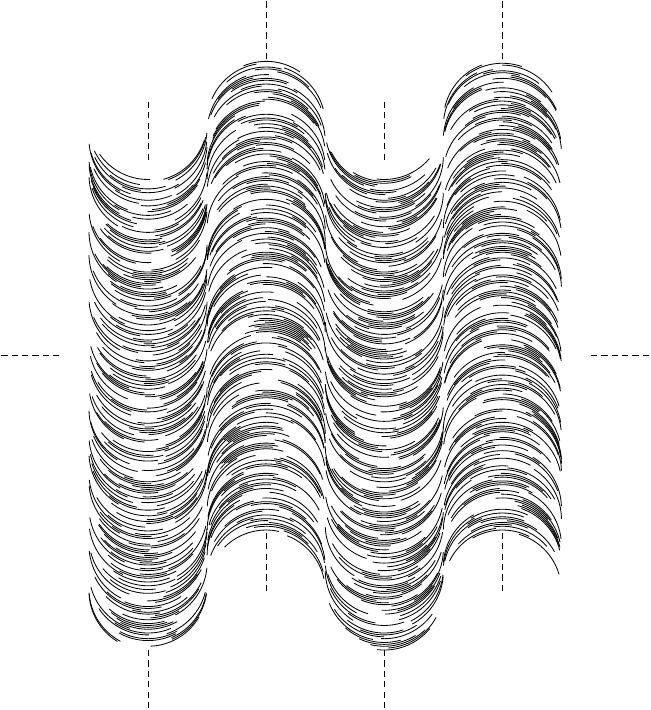}
\caption{Schematic illustration of a configuration of
an idealised prototypical version of the filamentary phase
in a system of hard infinitesimally--thin minor circular arcs
with, e.g.,  $\uptheta=1$. 
}
\label{lafigura3}
\end{figure}

In analogy with the phase behavior of hard spherical caps with $\uptheta \in [0,\pi]$ \cite[][(b)]{calotta},
systems of hard infinitesimally--thin minor circular arcs may form
a filamentary phase 
(Fig. \ref{lafigura3}).
In a filament of this phase, 
the hard infinitesimally--thin minor circular arcs tend
to organise on the same semicircumference with
the centres of the parent circles that ensuingly and randomly file;
a single filament is thus polar.
Different filaments of this phase may dispose themselves in a row 
along a direction approximately perpendicular to the filament axis, 
separated by a distance approximately equal to $2R$ and 
(anti)parallel oriented to adjacent filaments;
the filamentary phase is (more probably) non-polar.
If particularly preceded by a (quasi)nematic phase,
the formation of this phase can be revealed by
a decrease in the values of $\mathtt{S}_2$. 
More 
generally,
its formation  can be revealed by
the appearance of oscillations in 
the ${\mathcal G}_{1,\perp}^{\hat{\mathbf{u}}}(r_{\perp})$ and  
a sequence of equi-spaced peaks in the $g(r)$ and $G(c)$.

\begin{figure}
\centering
\includegraphics[scale=2.5]{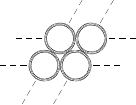}
\caption{
Schematic illustration of a configuration 
of an idealised prototypical version of a cluster hexagonal phase
in a system of hard infinitesimally--thin major circular arcs
with, e.g., $\uptheta=1.3 \uppi = 4.084 \cdots$.
Roundish clusters of 
hard infinitesimally--thin major circular arcs 
dispose themselves with their centres of mass at 
the sites of a triangular lattice. 
The hard infinitesimally--thin major circular arcs are with, e.g., $\uptheta=1.3 \uppi = 4.084 \cdots$ so that 
each roundish cluster is composed of a maximum of
$\mathtt{n}=6$ of them (Eq. \ref{equxn}, \cite{spiega}). 
The hard infinitesimally--thin major circular arcs are 
purposedly structurally organised in 
an expanded configuration 
to aid the appreciation of their (anti)clockwise ``vortical'' 
structural organisation.
Towards the densest--known packings,
each roundish cluster progressively contracts and 
the triangular--lattice spacing consequently decreases up
to a point that 
hard infinitesimally--thin major circular arcs essentially are on 
the same circumference and 
the triangular--lattice spacing is equal to 
$2R$.
}
\label{lafigura4}
\end{figure}

The structure of the densest--known packings of 
hard infinitesimally--thin major arcs \cite{hcadkp} 
suggests 
a suitably modified \textsl{hexatic bond--orientational} order parameter.
These densest--known packings and the corresponding
cluster hexagonal phase have a two--level structural organisation (Fig. \ref{lafigura4}). 
On the first level, 
a maximum of $\mathtt{n}$ (Eq. \ref{equxn}, \cite{spiega})  
hard infinitesimally--thin major circular arcs form 
roundish clusters 
that 
remind a vortex.
On the second level, 
these roundish clusters organise 
in configurations that 
remind the densest configuration of hard circles 
\cite{discretegeometry}. 
This second-level structural organisation  
suggests 
a hexatic bond-orientational order parameter $\uppsi_6$.
It is defined as
\begin{equation}
\uppsi_6 = \left \langle \frac{1}{\mathcal N} 
\left | \sum_{i=1}^{\mathcal N}
\frac{1}{{\rm{n}_{\rm vic}}_i} \sum_{j=1}^{{\rm{n}_{\rm vic}}_i } 
\mathsf{e}^{\mathtt{i} 6 \upvarphi_{ij} } \right |
\right \rangle 
\label{psi6eq}
\end{equation}
with: $\mathcal N$ the number of roundish clusters;
${\rm{n}_{\rm vic}}_i$ 
the number of vicinal roundish clusters $j$ of 
a certain roundish cluster $i$, 
defined as 
those roundish clusters $j$ whose 
centres of mass are 
within a pre-fixed distance from 
the centre of mass of 
the roundish cluster $i$;
$\upvarphi_{ij}$ the angle that the fictious ``bond''  
between the roundish clusters $i$ and $j$ 
forms with 
an arbitrary fixed axis.
The application of this order parameter naturally presupposes that
sufficiently compact and numerous roundish clusters are 
at least incipient.
This can be detected by $G(c)$ via a peak at $c=0$. 
Further growth of this peak 
together with the growth of the peak at $c=2R$ and 
the progressive split of the peak at $c \simeq 4R$ 
reveal that 
the processes of formation of roundish clusters and of
their hexagonal structural organisation are consolidating.

\subsection{monte carlo numerical simulations}
\label{mc}
Systems of hard infinitesimally-thin circular arcs
were investigated by 
Monte Carlo (MC) \cite{MCorigin,MClibri} method in 
the isobaric(--isothermal) ($NPT$) \cite{MCWood,MClibri} ensemble.
The number of particles usually was $N=600$,
although larger values of $N$ were also considered 
such as $N = 5400$ for various values of $\uptheta$ and 
occasionally $N=6400$ in the limit $\uptheta \rightarrow 0$. 
The $N$ hard infinitesimally--thin circular arcs
were placed in an
either rectangular or parallelogrammatic variable container.
The usual periodic boundary conditions were applied.       
The pressure $P$ was measured in units 
$k_B \, T \, \ell^{-2}$ with 
$k_B$ the Boltzmann constant, 
$T$ the absolute (thermodynamic) temperature and 
$\ell=\uptheta R$ the length of a circular arc.
For any value of $\uptheta$ that was investigated,
many values of the dimensionless pressure $P^{\star} = P \ell^2 /(k_B T)$ 
were considered.
For any value of these, 
the initial configuration was: 
Either a (dis)ordered configuration that was ad hoc constructed;
or a configuration that was previously
generated in a MC calculation at a nearby
value of $\uptheta$ or $P^{\star}$.
From the initial configuration,
the MC calculations (sequentially) proceeded.
Successive changes were attempted. 
Each of them was
randomly chosen among $2N+1$ possibilities:
With probability $N/(2N+1)$,
a random translation of the centroid of
a randomly selected particle;
with probability $N/(2N+1)$, 
a random rotation of the symmetry axis of 
a randomly selected particle;
with probability $1/(2N+1)$, 
a modification of one randomly selected side
of the container.
The (pseudo)random number generator that was employed was one that 
implements the Mersenne twister \texttt{mt19937} algorithm \cite{mt19937}.
These changes were accepted if no overlap resulted or rejected otherwise.
The acceptance of a change in 
the shape and size of the container was further subject to 
the ``Metropolis--like'' criterion that 
characterises the MC method in the $NPT$ ensemble 
\cite{MCWood,MClibri}.
For any specific values of $\uptheta$ and $P^{\star}$,
the maximal amounts of change were adjusted so that
20--30\% of each type of change could be accepted;
these adjustments were carried out 
in the course of exploratory 
MC calculations; 
the maximal amounts of change were not altered
in the course of subsequent MC calculations
that were conducted at those specific values
of $\uptheta$ and $P^{\star}$.
To improve on the efficiency of the MC calculations,
neighbour lists or linked-cell lists were employed \cite[particularly (a)]{MClibri}.
In both cases, the operative parameter was $r_{\rm cut} = 4 R \sin \left (\uptheta/4  \right)$ which is
the minimal distance at which 
two hard infinitesimally--thin circular arcs do not 
overlap 
irrespective of their mutual orientation. 
In the case of neighbour lists,
the list of neighbours of a particle $i$ comprised 
those particles $j$ whose distance from the centroid of $i$ was 
smaller than $r_{\rm cut} + r_{\rm skin}$; 
$r_{\rm skin}$ was that distance that had been selected 
in the course of exploratory MC calculations as 
the one that provided the largest efficiency.
Neighbour lists were automatically updated as soon as
$[r_{\rm skin}-2{\rm d}_{\rm max} ] \upkappa < r_{\rm cut}(1-\upkappa)$;
d$_{\rm max}$ was the maximum among the particle displacements
since the last update of the neighbour lists; 
$\upkappa$ was the ratio between
the new and old values of the length of the modified side.
In the case of linked-cell lists,
generally rectangular cells were constructed 
whose  minor side was at least equal to $r_{\rm cut}$ so that
the largest possible number of cells could be obtained.
Linked-cell lists were automatically updated as soon as,
following a change in the side of the  container:
Either the minor side of a cell 
became smaller than $r_{\rm cut}$ and 
thus a smaller number of cells had to be considered;
or it became sufficiently larger than $r_{\rm cut}$ to allow for 
more cells to be considered.
For any specific values of $\uptheta$ and $P^{\star}$,
exploratory MC calculations were conducted 
to decide
which type of lists led to the largest efficiency;
neighbour (linked-cell) lists were usually more efficient at higher (lower) density,
where the particle mobility was relatively small (large). 
It was also attempted 
to combine neighbour lists with linked-cell lists but 
to no avail: 
Efficiency did not significantly improve
with respect to 
separately considering the sole neighbour lists or linked-cell lists.    
The MC calculations were organised in cycles,
each of these comprising $2N+1$ attempts of a change.
For any specific values of $\uptheta$ and $P^{\star}$,
the MC calculations were  
subdivided into an equilibration
run and a production run.
Usually, an equilibration run lasted $10^7$ cycles while
the subsequent production run lasted as many cycles.
In the course of the production runs,
one every $10^4$ configurations was saved and
stored for the subsequent statistical analysis.
This statistical analysis comprised: 
The calculation of 
the mean number density $\left \langle \uprho \right \rangle$,
$\uprho$ being measured in units $\ell^{-2}$ so that
the dimensionless number density is 
$\uprho^{\star}=\uprho \ell^2$ and its mean
$\left \langle \uprho^{\star} \right \rangle = \left \langle \uprho \right \rangle \ell^2$; 
the calculation of 
the order parameters and correlation functions that 
are described in Section \ref{opcf}; 
the errors in $\left \langle \uprho^{\star} \right \rangle$ and 
in the order parameters were estimated by 
a habitual blocking method \cite{blocks}.

\section{results}
\label{phasediagram}

\subsection{description}

\begin{figure}
\includegraphics[scale=0.9]{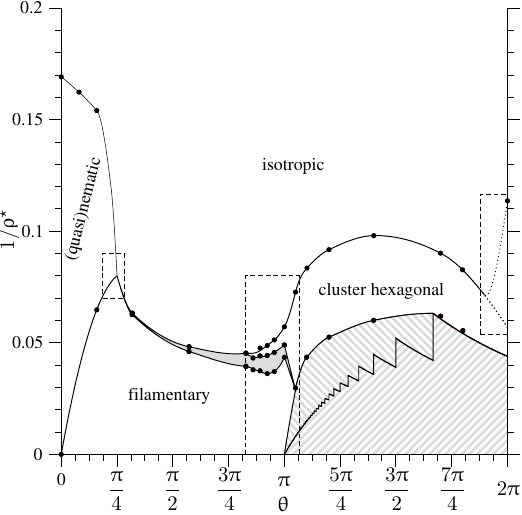}
\caption{
Phase diagram of systems of hard infinitesimally--thin circular arcs in the plane
aperture angle $\uptheta$ versus inverse of dimensionless number density $1/\uprho^{\star}$.
Black circles correspond to the original data that have been acquired from the MC numerical simulations while 
the solid lines that transverse them are guides to the eye.
The gray region on the left is the co-existence region that separates 
the filamentary phase from either the isotropic phase or the cluster hexagonal phase.
The hatched regions on the right are those regions that are effectively or theoretically prohibited as 
they correspond to values of number density that are 
higher than the values of number density of the effective or theoretical densest(--known) packings:
The upper mostly monotonic curve that 
transverses the relevant black circles and 
delimits the extra region that is hatched with oblique lines from top-left to bottom-right 
corresponds to 
the effectively densest packings whose number density has been acquired from the MC numerical simulations;
the lower zigzagging curve that 
delimits the region that is hatched with oblique lines from bottom-left to top-right 
corresponds to the theoretical densest--known packings \cite{hcadkp,spiega}.
The two dotted lines towards the hard--circle limits indicate 
the two possible scenarios of the isotropic--cluster hexagonal phase transition line while approaching that limit.
The three dashed rectangles enclose 
the most delicate regions of the phase diagram. 
}
\label{lafiguradiafase}
\end{figure}

By combining the equation of state and 
the set of order parameters and correlations functions (Section \ref{opcf}) and
with the aid of the visual inspection of configurations,
four (distinctive) phases have been identified. 
On varying $\uprho$ and $\uptheta$,
in addition to 
(I) a (quasi)nematic phase,
systems of hard infinitesimally--thin circular arcs can form:
(II) a (cluster) isotropic phase 
where, if $\uprho$ is sufficiently high, 
either filamentous or roundish clusters of 
hard infinitesimally--thin circular arcs are recognisable;
(III) a filamentary phase as schematically depicted as in Fig. \ref{lafigura3};
(IV)  a cluster hexagonal phase as schematically depicted as in Fig. \ref{lafigura4}.
The regions that the four phases occupy in 
the $\uptheta$ versus $1/{\uprho^{\star}}$ plane,
together with the curves that delimit them,
configure the phase diagram in Fig. \ref{lafiguradiafase}.
In describing this phase diagram,
it is convenient to subdivide it into four, $\uptheta$-dependent, sections:
({\sc i})   $ \displaystyle 0 \le \uptheta \lesssim \frac{\uppi}{4}$;
({\sc ii})  $ \displaystyle \frac{\uppi}{4} \lesssim \uptheta < \uppi$;
({\sc iii}) $ \uptheta \sim \uppi$;
({\sc iv})  $ \uptheta > \uppi$.

\subsubsection{$ 0 \displaystyle \le \uptheta \lesssim \frac{\uppi}{4}$}

The left-handed side of this section corresponds to 
the phase behavior of systems of hard segments.
The numerical simulation data for this basic reference system were usually interpreted as
inconsistent with a second--order isotropic--nematic phase transition that
the application of a second--virial (Onsager \cite{onsager}) 
density functional theory would predict \cite{kayser}. 
They were usually interpreted as 
consistent with the existence of an isotropic--(quasi)nematic phase transition of 
the Berezinskii--Kosterlitz-Thouless \cite{berezinskii,kt,ktrev,ktrus} type 
\cite{eppenga,india,vink}. 
One interpretation that essentially maintains both of these two, 
usually mutually exclusive, interpretations was also proposed \cite{polonia}.
In an infinite periodic system, 
the \textit{S}-shaped curve of $\mathtt{S}_2$ versus $\uprho$ is suggestive of 
an isotropic--nematic phase transition. 
In three dimensions, it would be indeed taken as a signature of such a phase transition.
In two dimensions, it is instead considered insufficient. 
This insufficiency is based on assuming that  
basic analytic results 
for specific two--dimensional systems \cite{mermin1,mermin2} 
have to also preclude a proper long--ranged nematic ordering
in a two-dimensional system.  
Even though those analytic results were found inapplicable to 
a two-dimensional nematic phase that is formed
in a realistic system of particles interacting via non--separable interactions 
as hard particles are \cite{straley}.
Based on that paradigm, 
$\mathtt{S}_2$ of an infinite (thermodynamic) system would be equal to zero at all values of $\uprho$.
For this reason, one should turn to 
explicitly considering ${\mathcal{G}}_2(r)$ and its long-distance behavior.
The latter distinguishes the two phases at either side of 
a phase transition of the Berezinskii--Kosterlitz-Thouless type:
In the isotropic phase, ${\mathcal{G}}_2(r)$ decays to zero exponentially;
in the (quasi)nematic phase, ${\mathcal{G}}_2(r)$ decays to zero algebraically.
Even though past and present numerical simulation data seem to be consistent with this scenario,
the limited size of the systems that are considered in 
these numerical simulations cannot afford to clearly and unambiguously discern 
the characteristics of the long-distance decay. 
It is difficult to extrapolate to a very long distance 
the behavior of a correlation function that is known up 
to a decade of distance units. 
Based on this modest distance interval, 
it is  difficult to affirm what is the best fitting function overall.
It seems that, for sufficiently large values of $\uprho$, 
an algebraic fitting function outperforms an exponential fitting function.
However, other fitting functions could perform even better:
e.g., a ``stretched-exponential'' function fares at least as well as an algebraic function.

Based on these considerations, the attitude of this work is very pragmatic. 
In analogy to previous works \cite{eppenga,india}, 
${\mathcal{G}}_2(r)$ has been fitted to either an exponential or algebraic function. 
The value of $\uprho$ at which the latter fitting function seems 
to outperform the former fitting function is taken as 
the value that delimits the isotropic phase and the (quasi)nematic phase.
This is done without claiming it as 
objectively supporting a phase transition of 
the Berezinskii--Kosterlitz-Thouless type while 
conceding the present impossibility to 
more profoundly investigate
the nature 
of the two-dimensional nematic phase and 
of 
the  transition 
that separates it from 
the isotropic phase. 

\begin{figure}
\centering
\includegraphics[trim=0.5cm 0.0cm 6cm 0cm,clip]{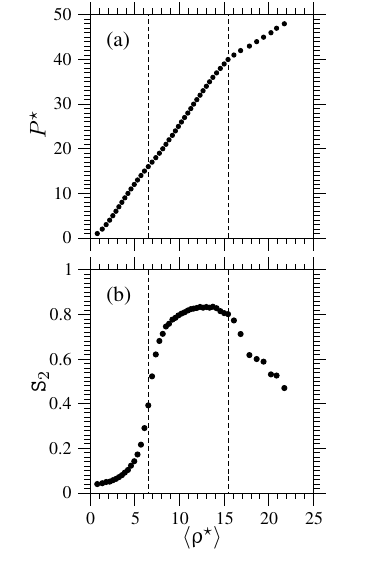}
\caption{
(a) equation of state $P^{\star}$ versus $\langle \uprho^{\star} \rangle$ and 
(b) nematic order parameter $\mathtt{S}_2$ versus $\langle \uprho^{\star} \rangle$ for 
a system of hard infinitesimally--thin circular arcs with $\uptheta=0.5$.
The dashed vertical line on the left separates the isotropic phase and the (quasi)nematic phase
while the dashed vertical line on the right separates the (quasi)nematic phase and the filamentary phase.
}
\label{figura1sezione1}
\end{figure}
\begin{figure}
\centering
\includegraphics[trim=0.5cm 1.5cm 4cm 0cm,clip]{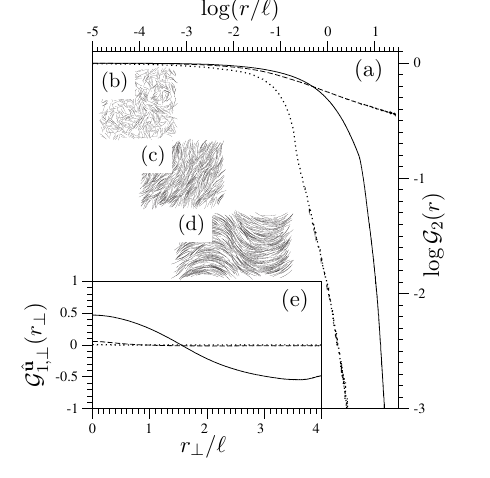}
\caption{(a) The orientational pair--correlation function ${\mathcal{G}}_2(r)$
in the isotropic phase at $P^{\star}=10$ (dotted line), (quasi)nematic phase
at $P^{\star}=35$ (dashed line) and filamentary phase at $P^{\star}=45$ (solid line).
Image of a configuration in the (b) isotropic phase at $P^{\star}=10$, (c) (quasi)nematic phase
at $P^{\star}=35$ and (d) filamentary phase at $P^{\star}=45$.
(e) The orientational pair--correlation function ${\mathcal G}_{1,\perp}^{\hat{\mathbf{u}}} (r_{\perp})$
in the isotropic phase at $P^{\star}=10$ (dotted line), (quasi)nematic phase
at $P^{\star}=35$ (dashed line) and filamentary phase at $P^{\star}=45$ (solid line). 
}
\label{figura2sezione1}
\end{figure}

In a system of hard infinitesimally--thin minor circular arcs with $\uptheta=0.25$, 
the isotropic and (quasi)nematic phases are
the sole phases that have been observed 
in the interval of values of $P$ that has been presently investigated.
Hard infinitesimally--thin minor circular arcs with $\uptheta=0.5$ are 
instead sufficiently curved for 
 another, denser and arguably more interesting, phase 
to succeed the (quasi)nematic phase already in 
the interval of values of $P$ that has been presently investigated.
This phase transition is revealed by 
a bent    in the equation of state and 
a descent in the values of $\mathtt{S}_2$ (Fig. \ref{figura1sezione1}).
These two signs   
are accompanied by 
a significant change in 
the long-distance behavior of ${\mathcal{G}}_2(r)$:
On going from the lower-density phase 
to the higher-density phase,
the long-distance behavior of 
${\mathcal{G}}_2(r)$
seems as if it revert to that in the isotropic phase:
No remnant of a possible algebraic decay remains 
[Fig. \ref{figura2sezione1} (a)]. 
One appreciates that 
the phase that spontaneously forms at larger values of $\uprho$ is 
the filamentary phase [Fig. \ref{figura2sezione1} (b, c, d)]. 
In fact, this phase is characterised by 
hard infinitesimally--thin minor circular arcs 
tending to organise on the same semicircumference; 
in turn, these generated semicircular clusters 
file to generate filaments;  
in turn, these filaments tend to mutually organise 
side-by-side and up-side-down [Fig. \ref{figura2sezione1} (d)].
Consistently, 
${\mathcal G}_{1,\perp}^{\hat{\mathbf{u}}} (r_{\perp})$
exhibits 
a short-distance oscillatory behavior with 
a period approximately equal to $2R$ [Fig. \ref{figura2sezione1} (e)].
It is conceivable that the filamentary phase also forms in 
systems of hard infinitesimally--thin minor circular arcs with $\uptheta < 0.5$
at increasingly higher density and pressure than 
those that have been presently investigated.  

By the concomitant action of both 
the isotropic phase at lower $\uprho$ and 
the filamentary phase at higher $\uprho$,
the number density interval in which 
the (quasi)nematic phase exists precipitously contracts 
as $\uptheta$ increases until 
this phase disappears  at $ \displaystyle \uptheta \simeq \frac{\uppi}{4}$.

\subsubsection{$ \displaystyle \frac{\uppi}{4} \lesssim \uptheta < \uppi$}

\begin{figure}
\centering
\includegraphics[trim=0.5cm 0cm 6cm 0cm,clip]{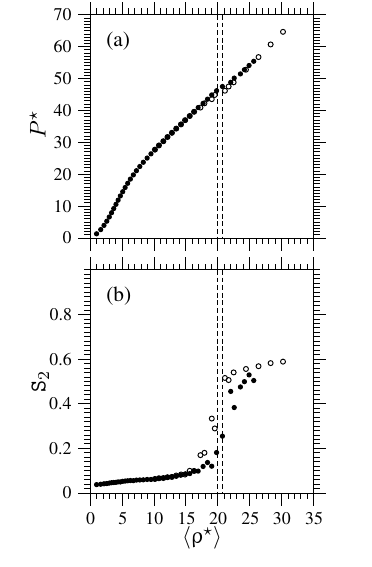}
\caption{(a) equation of state $P^{\star}$ versus $\langle \uprho^{\star} \rangle$ 
and (b) nematic order parameter $\mathtt{S}_2$ versus $\langle \uprho^{\star} \rangle$
for a system of hard infinitesimally--thin circular arcs with $\uptheta=1.8$.
In both panels, filled circles correspond to data that were obtained
by progressively compressing the system from 
an initial dilute and disordered configuration 
while empty circles correspond to data that were obtained by progressively
decompressing the system from an initial filamentary configuration
as that schematically depicted as in Fig. \ref{lafigura3}.
The two dashed vertical lines separate the isotropic phase and the filamentary phase.}
\label{figura1sezione2}
\end{figure}

\begin{figure}
\centering
\includegraphics[trim=0.5cm 1.5cm 4cm 0cm,clip]{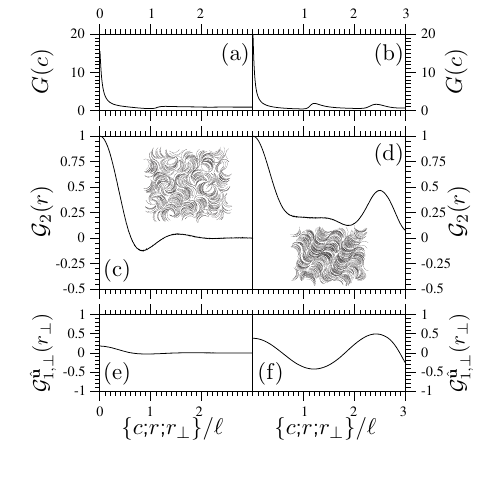}
\caption{For a system of hard infinitesimally--thin
circular arcs with $\uptheta=1.8$ in the isotropic phase at $P^{\star} = 40$ and 
in the filamentary phase at $P^{\star} = 53$,
the positional pair-correlation function $G(c)$ (a and b, respectively), 
the orientational pair-correlation function ${\mathcal{G}}_2(r)$ 
(c and d, respectively) and
the orientational pair-correlation function 
${\mathcal G}_{1,\perp}^{\hat{\mathbf{u}}} (r_{\perp})$ (e and f, respectively)
are shown.
The inset in (c) is an image of a configuration in the isotropic phase at $P^{\star} = 40$ while 
the inset in (d) is an image of a configuration in the filamentary phase at $P^{\star} = 53$.}
\label{figura2sezione2}
\end{figure}

In this section, 
the isotropic and filamentary are 
the sole phases that have been observed.
These two phases are separated by a first-order phase transition 
whose strength increases with increasing $\uptheta$.
This is revealed by the behavior of the equation of state 
[Fig. \ref{figura1sezione2} (a)].
$\mathtt{S}_2$ concurs to reveal this phase transition:
$\mathtt{S}_2$ exhibits a surge in correspondence to
the values of $\uprho$ at which the phase transition occurs;
the values that this order parameter takes on 
in the filamentary phase are significantly smaller than 
those typical of a (quasi)nematic phase [Fig. \ref{figura1sezione2} (b)].
In fact, 
in an idealised prototypical filamentary phase 
as schematically depicted as in Fig. \ref{lafigura3},
the  $\mathtt{S}_2$ would take on 
a value equal to $\displaystyle \frac{\sin(\uptheta-\uppi)}{\uptheta-\uppi}$.
The structural differences that occur on going from the isotropic phase to
the filamentary phase are revealed by the various pair--correlation functions 
(Fig. \ref{figura2sezione2}). 
Particularly, $G(c)$ becomes to peak at $c=2R$ and $c=4R$ [cf. (a) and (b) in Fig. \ref{figura2sezione2}],
while ${\mathcal G}_{1,\perp}^{\hat{\mathbf{u}}} (r_{\perp})$ becomes oscillatory
with a period equal to $4R$ [cf. (e) and (f) in Fig. \ref{figura2sezione2}]. 
On increasing $\uptheta$,
as the transition to the filamentary phase is approached,
the isotropic phase passes  
from being ordinary to exhibiting clusters.
These clusters are made of 
hard infinitesimally--thin circular arcs that
tend to organise on the same semicircumference and 
then file to generate filaments that are of 
varying length, degree of ramification and tortuousity [inset in Fig. \ref{figura2sezione2} (c)].
The progressive straightening of the equation of state is 
a symptom of the formation of these ``supraparticular'' structures that 
precurse a proper filamentary phase.
On increasing $\uptheta$,
in the same filamentary phase, 
the filaments tend to be more tortuous and 
it is increasingly more frequent to observe 
ramifications and ``ruptures''. 
These ramifications and ``ruptures'' are provoked by 
hard infinitesimally--thin circular arcs that tend to 
dispose in an antiparallel configuration [inset in Fig. \ref{figura2sezione2} (d)].

\subsubsection{$ \uptheta \sim \uppi$}

\begin{figure}
\centering
\includegraphics[trim=0.5cm 0cm 6cm 0cm,clip]{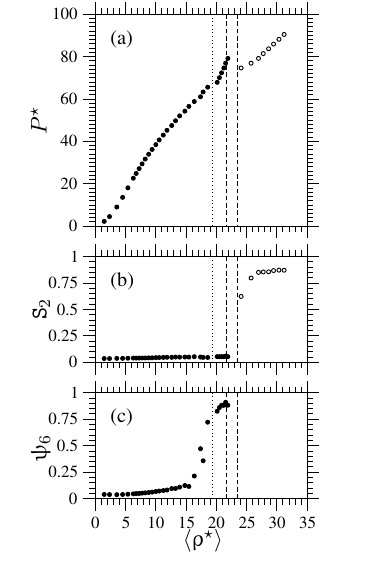}
\caption{
(a) equation of state $P^{\star}$ versus $\langle \uprho^{\star} \rangle$,
(b) nematic order parameter $\mathtt{S}_2$ versus $\langle \uprho^{\star} \rangle$ and 
(c) hexatic order parameter $\uppsi_6$ versus $\langle \uprho^{\star} \rangle$
for a system of hard infinitesimally--thin circular arcs with $\uptheta=3$.
In panels (a) and (b) 
filled circles correspond to data that were obtained 
by progressively compressing the system from 
an initial dilute and disordered configuration 
while empty circles correspond to data that were obtained 
by progressively decompressing the system from 
an initial filamentary configuration
as that schematically depicted as in Fig. \ref{lafigura3}.
In the three panels, 
the vertical dotted line separates 
the isotropic phase and the cluster hexagonal phase while 
the two vertical dashed lines separate 
the cluster hexagonal phase and the filamentary phase.
}
\label{figura1sezione3}
\end{figure}

In this section, a new, arguably most interesting, phase appears 
in between the isotropic phase and the filamentary phase: 
The cluster hexagonal phase.
In the isotropic phase, 
the tendency that filamentous clusters have to break and close up 
increases up to conducing to the formation of roundish clusters.
This occurs up to a point that 
the roundish clusters become sufficiently compact and numerous 
and 
their number sufficiently large
to organise in
a triangular lattice.
The formation of this cluster hexagonal phase, 
which prevents the spontaneous formation of the filamentary phase, 
can be revealed by 
examining the equation of state:  
It corresponds to a tenuous surge in its graph that 
is recognisable at 
values of $\uprho^{\star} \simeq 20$ [Fig. \ref{figura1sezione3} (a)].
While 
$\mathtt{S}_2$ is unable to reveal this phase transition [Fig. \ref{figura1sezione3} (b)],
better evidence of a transition between 
the isotropic phase and the cluster hexagonal phase is nonetheless acquired by
examining the dependence of 
$\uppsi_6$ 
on 
$\uprho$:  
This order parameter 
exhibits a clear surge in correspondence to
the isotropic--cluster hexagonal phase transition 
[Fig. \ref{figura1sezione3} (c)].
$\mathtt{S}_2$ can instead
distinguish between the cluster hexagonal phase and the filamentary phase: 
Since the roundish clusters are overall isotropic,
$\mathtt{S}_2$ (effectively) vanishes in the cluster hexagonal phase 
as it does in the isotropic phase;
since the structural units of the filamentary phase
are formed by a progressively smaller number of
hard infinitesimally--thin minor circular arcs as 
$\uptheta \rightarrow \uppi$,
$\mathtt{S}_2$ is increasingly significantly larger than zero
in the filamentary phase 
 [Fig. \ref{figura1sezione3} (b)].
The two cluster phases are separated by a first--order
phase transition. 
\begin{figure}
\centering
\includegraphics[trim=0.5cm 1cm 4cm 0cm,clip]{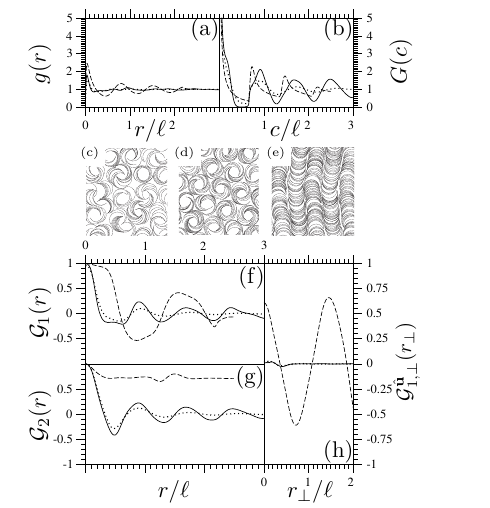}
\caption{
For a system of hard infinitesimally--thin circular arcs with $\uptheta = 3$ 
in the isotropic phase at $P^{\star} = 25$, the cluster hexagonal phase at $P^{\star} = 34$ and the filamentary phase at $P^{\star} = 34$:
(a) the positional pair--correlation function $g(r)$;
(b) the positional pair--correlation function $G(c)$;
(c, d, e) image of a configuration respectively in the isotropic phase, cluster hexagonal phase and filamentary phase;
(f) the orientational pair--correlation function ${\mathcal{G}}_1(r)$;
(g) the orientational pair--correlation function ${\mathcal{G}}_2(r)$;
(h) the orientational pair--correlation function ${\mathcal G}_{1,\perp}^{\hat{\mathbf{u}}} (r_{\perp})$;
in panel (a, b, f, g, h), the dotted, solid and dashed lines are respectively for the isotropic, cluster hexagonal and filamentary phases.
}
\label{figura2sezione3}
\end{figure}
The structural differences among the three phases are revealed by 
the various pair--correlation functions and evidenced by 
the corresponding images of a configuration (Fig. \ref{figura2sezione3}).
Based on these images, 
one notes the similarity between 
the structures of the isotropic phase and of the cluster hexagonal phase which 
contrast with 
the structure of the filamentary phase.
This (dis)similarity among the three phases is reflected in 
the graphs of the various pair--correlation functions (Fig. \ref{figura2sezione3}).

\subsubsection{$ \uptheta > \uppi$}


\begin{figure}
\centering
\includegraphics[trim=1cm 4cm 2.5cm 2.5cm,clip]{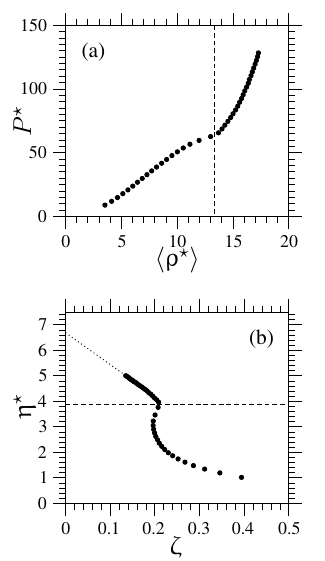}
\caption{
equation of state  
(a) in the representation  $P^{\star}$ versus $\langle \uprho^{\star} \rangle$ and 
(b) in the representation $\upeta^{\star}$ versus $\upzeta$ for 
a system of hard infinitesimally--thin circular arcs with $\uptheta=1.1 \uppi = 3.455 \cdots$;
in (a) the vertical dashed line and 
in (b) the horizontal dashed line separates
the isotropic phase and the cluster hexagonal phase; 
in (b) the dotted line is a linear fit extrapolation of 
the higher-density part of the solid-phase branch.
}
\label{figura1sezione4_A}
\end{figure}

\begin{figure}
\centering
\includegraphics[trim=0.5cm 0cm 0cm 0cm,clip]{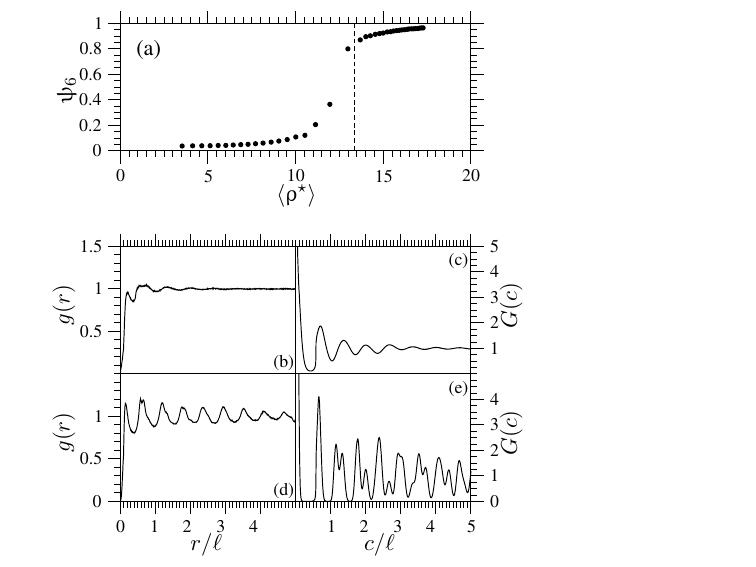}
\caption{
For a system of hard infinitesimally--thin circular arcs with $\uptheta=1.1 \uppi = 3.455 \cdots$:
(a) the order parameter $\uppsi_6$ as a function of number density $\langle  \uprho^{\star} \rangle$;
the vertical dashed line separates the isotropic phase and the cluster hexagonal phase;
the pair--correlation functions $g(r)$ and $G(c)$ 
in the isotropic phase at $P^{\star}=56.7$ (b, c) and 
in the cluster hexagonal phase at $P^{\star}=128.4$ (d, e). 
}
\label{figura2sezione4}
\end{figure}

The fact that $\uptheta$ surpasses the intermediate value of $\uppi$
is very consequential.
It was already observed that 
two hard infinitesimally--thin major circular arcs that 
are disposed on top of one another
cannot superpose; 
they can only superpose if 
they are suitably rotated with respect to one another in a way that, 
once it is exactly replicated 
$\mathtt{n}$ (Eq. \ref{equxn}, \cite{spiega}) - 2 times,
conduces to the formation of 
those compact circular clusters that characterise 
the corresponding densest--known packings \cite{hcadkp}. 
This fact significantly destabilises the filamentary phase
with respect to 
the cluster hexagonal phase:  
The former phase precipitously disappears
leaving the latter phase as 
the sole observable phase at sufficiently high $\uprho$.
The cluster hexagonal phase is separated from the isotropic phase by 
a transition whose weakness presently makes impossible 
to assess whether 
it is either (more probably) first-order or second-order.
This phase transition is revealed by 
a visually recognisable change in 
the graph of the equation of state
[Fig. \ref{figura1sezione4_A} (a)].
This change may be made clearer by 
plotting the effective packing fraction 
$\displaystyle \upeta^{\star} = 
\frac{\uprho \uppi R^2}{\uppi/(2\sqrt{3})}$
with respect to 
the inverse compressibility factor 
$\upzeta = \displaystyle \frac{\uprho^{\star}}{P^{\star}}$ 
[Fig. \ref{figura1sezione4_A} (b)].
One further revealer of 
the formation of a cluster hexagonal phase is again
$\uppsi_6$: 
It exhibits a surge in correspondence to 
$\langle \uprho^{\star} \rangle \simeq 12$,
the value of $\langle  \uprho^{\star} \rangle$ at which
the isotropic--cluster-hexagonal phase transition occurs 
[Fig. \ref{figura2sezione4} (a)]. 
In addition, 
the form of 
$g(r)$ and $G(c)$ passes from being fluid--like 
[Fig. \ref{figura2sezione4} (b, c)]
 to being
crystalline--like 
[Fig. \ref{figura2sezione4} (d, e)].

\begin{figure}
\centering
\includegraphics[trim=0.5cm 4cm 2.5cm 2.5cm,clip]{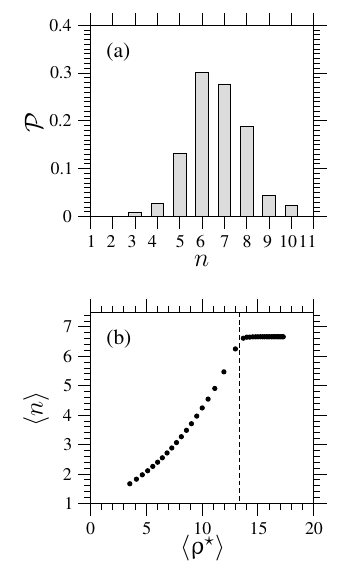}
\caption{
(a) Probability distribution function $\mathcal P$ of 
the number $n$ of 
hard infinitesimally--thin circular arcs per roundish cluster for 
a system of hard infinitesimally--thin circular arcs with $\uptheta=1.1 \uppi = 3.455 \cdots$ at $P^{\star}=128.4$.
(b) The mean value $\langle n \rangle$ of the number of hard infinitesimally--thin circular arcs per
roundish cluster as a function of $\langle \uprho^{\star} \rangle$ 
for a system of hard infinitesimally--thin circular arcs with $\uptheta=1.1 \uppi = 3.455 \cdots$;
the vertical dashed line separates the isotropic phase and the cluster hexagonal phase. 
}
\label{figura1sezione4_B}
\end{figure}

\begin{figure}
\centering
\includegraphics[trim=1cm 1.25cm 1cm 1cm,clip]{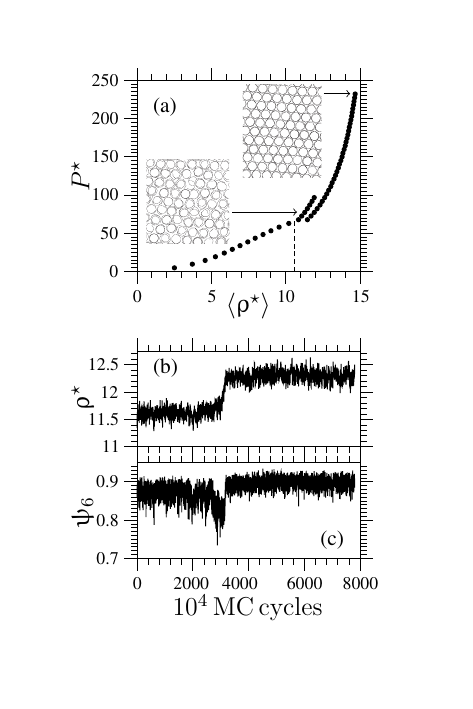}
\caption{For a system of hard infinitesimally--thin circular arcs with $\uptheta = 1.4 \uppi = 4.398 \cdots$: 
(a) equation of state $P^{\star}$ versus $\langle \uprho^{\star} \rangle$; 
note that there are two cluster-hexagonal branches; 
the vertical dashed line separates the isotropic phase and the lower-density cluster-hexagonal phase;
the inset on the bottom left is an image of a configuration taken at $P^{\star} = 77.4$ 
in the lower-density cluster-hexagonal branch;
the inset on the top right is an image of a configuration taken at $P^{\star} = 232.1$
in the higher-density cluster-hexagonal branch;
(b) evolution of the number density $\uprho^{\star}$ at $P^{\star} = 77.4$ as a function of MC cycles; 
(c) evolution of the number density $\uppsi_6$ at $P^{\star} = 77.4$ as a function of MC cycles.
}
\label{figura1sezione4_C}
\end{figure}

In the graphical representation of Fig. \ref{figura1sezione4_A}(b), 
one can observe that
the cluster-hexagonal-phase branch is, to a good approximation, linear.
This is consistent with the applicability, also to the present case, of 
a suitably adapted version of 
the free--volume theory \cite{fvt1}. 
This theory is known to provide 
a good (the) description and interpretation of 
the equation of state of 
a dense solid phase in 
a system of hard particles \cite{fvt1,fvt2}.
In an equilibrium system of hard circles (discs),
the linear extrapolation of the high-density solid-branch curve 
would intersect the ordinate axis at a value equal to 1.
The linear extrapolation of 
the high-density solid-branch curve 
in a system of hard infinitesimally--thin major circular arcs with $\uptheta = 1.1 \uppi = 3.455 \cdots$ 
intersects the ordinate axis at a value approximately equal to 6.5 
[Fig. \ref{figura1sezione4_A} (b)]. 
This value corresponds to 
the mean value $\langle n \rangle$ of 
the hard infinitesimally--thin major circular arcs with 
$\uptheta = 1.1 \uppi = 3.455 \cdots$ per roundish cluster.
This is confirmed by 
a more direct calculation of 
$\langle n \rangle$. 
It ensues from the calculation of 
the probability distribution, $\mathcal P$, of the number, $n$, of 
hard infinitesimally--thin major circular arcs per roundish cluster: 
${\mathcal P}(n)$ [Fig. \ref{figura1sezione4_B} (a)].
The value of $\langle n \rangle$ increases with 
$\langle \uprho ^{\star} \rangle$ in  
the isotropic phase until  
it flatly levels up as 
the system enters the cluster hexagonal phase 
[Fig. \ref{figura1sezione4_B} (b)].
The limit value $\langle n \rangle \simeq 6.5$ in
the cluster hexagonal phase is smaller than 
${\mathtt{n}} = 19 $ 
(Eq. \ref{equxn}, \cite{spiega})
\cite{hcadkp}.
This means that the cluster hexagonal phase that 
spontaneously forms from the isotropic phase is 
a sub--optimal version of 
these densest--known packings.
This is comprehensible as 
the phase transition occurs at 
a value of $\langle \uprho ^{\star} \rangle \simeq 12$ that 
is still relatively small [Fig. \ref{figura1sezione4_B} (b)].
Yet, the subsequent constancy of 
$\langle n \rangle$ with $\langle \uprho ^{\star} \rangle$  
[Fig. \ref{figura1sezione4_B} (b)]
raises two questions as to 
whether the densest--known packings could ever spontaneously form 
on progressive compression and 
whether the cluster hexagonal phase that 
spontaneously forms from the isotropic phase could ever be 
an equilibrium phase.
The importance and relevance of these questions also emerge from 
investigating  systems of 
hard infinitesimally--thin major circular arcs 
with a larger value of $\uptheta$.
In these cases, 
more than one cluster hexagonal phase branches can be observed; 
these  are separated by what would seem as 
a first--order phase transition [Fig. \ref{figura1sezione4_C} (a)]. 
Such a discontinuous phase behavior can be also appreciated by
examining the evolution of $\uprho^{\star}$ in the course of 
a Monte Carlo numerical simulation:
An abrupt jump in the values of $\uprho^{\star}$ is frequently observed 
[Fig. \ref{figura1sezione4_C} (b)].
This is due to the roundish clusters that 
are progressively re-organising in such a way that 
they incorporate, on the average, 
more constituting hard infinitesimally--thin circular arcs:
In correspondence to the rise in the values of $\uprho^{\star}$ 
there is a momentary fall in the values of $\uppsi_6$;
this suggests a momentary re-organisation of the system that 
allows the incorporation of more 
hard infinitesimally--thin circular arcs into
the same roundish cluster(s) [Fig. \ref{figura1sezione4_C} (c)].
The lower-density branch directly forms from 
the isotropic phase [Fig. \ref{figura1sezione4_C} (a)].
If pressure is sufficiently high, 
then it transforms into 
the higher-density branch 
after many MC cycles (Fig. \ref{figura1sezione4_C}).
From that value of pressure, 
the higher-density branch can be continued 
up to higher pressure and 
down to lower pressure 
[Fig. \ref{figura1sezione4_C} (a)].  
One may assess the lower-density branch in 
Fig. \ref{figura1sezione4_C} (a) 
as an enduring metastable branch.
However, 
the same assessment is applicable to 
the unique branch that is observed in Fig. \ref{figura1sezione4_A} and 
to the higher-density branch in Fig. \ref{figura1sezione4_C} (a) 
with respect to 
other, hypothesisable, even higher-density branches that 
insufficiently lengthy MC numerical simulations prevent from observing.
Continuous incorporation and release of 
hard infinitesimally--thin circular arcs
into or from roundish clusters
are necessary to spontaneously attain and maintain 
an equilibrium ${\mathcal P}(n)$.
In a cluster hexagonal phase,
it is conceivable that 
the mechanisms of incorporation and release become 
increasingly less effective as $\uprho$ increases.
Based on Figs. 
\ref{figura1sezione4_A}, 
\ref{figura1sezione4_B}, 
\ref{figura1sezione4_C},
it is presently unclear how   
an equilibrium cluster hexagonal phase should proceed towards 
the densest--known packing limit:
Either continuously, 
via a gradual modification of ${\mathcal P}(n)$;
or discontinuously, 
via a sequence of iso--structural first--order phase transitions,
each phase at either side of the phase transition being 
a cluster hexagonal phase
with its own ${\mathcal P}(n)$.
The discontinuous behavior that is observed may well be 
an artificial effect due to 
the finite size of  the systems that are considered in 
the MC numerical simulations 
which is further exacerbated by 
an (always looming) insufficiency of their duration.

On progressive compression from the isotropic phase,
the capability of hard infinitesimally--thin major circular arcs of
intertwining in roundish clusters 
persists up to values of $\uptheta$ almost equal to $2 \uppi$:
For a value of $\uptheta$ as large as $1.9 \uppi = 5.969 \cdots$ 
a vast majority of dimers are observed.
This capability should deteriorate as 
the value of $\uptheta$ is further increased:
in the very close neighbourhood of $\uptheta=2 \uppi$, 
a more even mixture of 
monomers and dimers is expected, 
with  the former 
progressively becoming more abundant as 
the hard-circle limit is approached. 
In the hard-circle limit, 
the system is clearly formed by single hard circles.
However, it is probable that 
this limit should be considered as 
a singular limit.
This hypothesis relies on 
the densest--known packings being formed by dimers that 
dispose themselves at the sites of a triangular lattice for 
any value of $\uptheta < 2 \uppi$ \cite{hcadkp}. 
In the thermodynamic limit at very high density, 
the stablest structures should correspond to 
these densest--known packings 
for any value of $\uptheta < 2 \uppi$ \cite{hcadkp}. 
 
\subsection{discussion}

In discussing the phase diagram that has been described,
there are several comparisons to be made and 
connections to be established 
with previous works.

Section ({\sc{i}}) resembles the phase diagram of 
a generalised planar rotor spin system on a square lattice. 
In this system, 
the interaction of a spin $i$ with 
its nearest neighbour spin $j$ comprises 
the (ferromagnetic or) polar term, 
proportional to 
$\cos\left ( \upvarphi_j - \upvarphi_i \right)$, and 
the (generalised) nematic term, 
proportional to 
$\cos\left ( q \left( \upvarphi_j - \upvarphi_i \right) \right)$,
with $q=\rm{2,3,4}$ \cite{planarot1985,planarot1989,planarot2011,planarot2012}.
This resemblance is made once
the low-density isotropic (high-density filamentary) phase 
in the present off--lattice systems 
is associated to 
the high-temperature paramagnetic (low-temperature ferromagnetic) phase 
in those on--lattice systems.
In both types of systems, 
a (quasi)nematic phase exists in between the other two respective phases.
Its stability interval diminishes as, respectively, 
either $\uptheta$ increases 
or the polar term proportional to $\cos\left ( \upvarphi_j - \upvarphi_i \right)$ prevails.
However, this resemblance may be solely qualitative and 
a possible correspondence between the two types of systems may immediately end.
Even though sharing the same rotational symmetry, 
investigations on on--lattice systems of spins interacting with 
a potential energy $\rm U$ of the form
\[
{\rm U} =  -a_1 \cos(\upvarphi_j -\upvarphi_i) - a_q \cos \left( q (\upvarphi_j - 
 \upvarphi_i)  \right)
\]
with $q \ge 2$ have shown that the resulting phase diagrams
can significantly depend on $q$ \cite{planarot2011}.
It is presumable that the same conclusion hold
when other, more general, expressions for the potential energy 
of the form
\[
{\rm U} = - \sum_{q \ge 1} a_q \cos \left( q (\upvarphi_j - 
 \upvarphi_i)  \right)
\] 
were considered \cite{zukovic}.
The excluded area could be considered as the quantity 
in a two--dimensional off--lattice non--thermal model
that plays a role analogous to 
the potential energy 
in a two--dimensional on--lattice thermal model.
In fact, there has been an investigation of 
a system of spins on a square lattice interacting with 
a potential energy  of the same form as 
the excluded area between two discorectangles \cite{pv}. 
If one adapted this exercise to the present case,
one would have to 
calculate the excluded area between two hard infinitesimally--thin
minor circular arcs.
The first four terms of 
the Fourier series expansion of this excluded area (Table \ref{tabella})
would indicate that 
the effect of progressively curving hard segments into 
hard infinitesimally--thin minor circular arcs
would be to generally make the effective interactions more isotropic.
\begin{table}
\begin{tabular}{|c|c|c|c|c|}
\hline
\hline
$\uptheta$ & $a_1/a_2$ & $a_2/a_2$ & $a_3/a_2$ & $a_4/a_2$ \\
\hline
\hline
0       &  0        & 1           &  0        & 0.2 \\
\hline
0.11285 & -0.000762 & 1           & -0.000754 & 0.198 \\
\hline
0.56608 & -0.0214   & 1           & -0.0196   & 0.158 \\
\hline
1.14391 & -0.107    & 1           & -0.0755   & 0.0873 \\
\hline
\hline
\end{tabular}
\caption{the first four coefficients of the Fourier series of the excluded
area between two hard infinitesimally--thin circular arcs that subtend an angle $\uptheta$;
note that the coefficients are in units of the coefficient $a_2$.}
\label{tabella}
\end{table}
These effective interactions would be essentially classifiable
as a composition of 
an (antiferromagnetic or) antipolar term with a nematic term 
(Table \ref{tabella}).
This would not be entirely compatible with 
the observation of 
a filamentary phase which, 
although globally non-polar, is locally polar.
The passage from an on--lattice model to an off--lattice model
does not seem to be that straightforward: 
If even maintaining the particles constrained on a lattice
and changing a sole term proportional to 
$\cos \left( q (\upvarphi_j -  \upvarphi_i) \right)$ 
produces qualitatively different phase diagrams,
then results that may be valid for an on--lattice system
may not apply, especially in low dimensions,
to a supposedly related off--lattice system.
Not only are the centroids of 
the hard infinitesimally--thin circular arcs 
not constrained on a lattice but 
the interaction between two of them 
is also non--separable.
This significantly adds to debilitating 
a possible association between 
the filamentary phase in 
a system of hard infinitesimally-thin minor circular arcs and 
the ferromagnetic phase in 
a system of generalised planar rotor spins:
In fact, 
even leaving aside the positional structure of the filamentary phase,
its orientational pair--correlation functions are 
significantly more complicated than 
the simple algebraically and monotonically decaying 
pair--correlation functions
in the ferromagnetic phase.  


The filamentary phase is the same phase that 
was observed in numerical simulations on 
systems of hard bow--shaped particles formed by three suitably disposed hard segments \cite[(a)]{zigzag}.
In that work, this phase was denoted as ``modulated nematic'' phase.
In a subsequent work that 
attempted to reproduce these numerical simulation data with 
second-virial (Onsager \cite{onsager}) density-functional theory analytic calculations, 
this phase was interpreted as 
a ``splay--bend nematic'' phase \cite[(b)]{zigzag}.
Irrespective of 
whether it could be qualified as generically ``modulated'' or particularly ``splay--bend'',
it is actually 
the application of the adjective ``nematic'' to this phase that does not convince.
One constitutive feature of a nematic phase
is its positional uniformity.
The filamentary phase, 
even in those versions that are very rippled with 
ramifications, ruptures and tortuousity, 
is not positionally uniform: 
e.g., the probability density to find a particle intra-filament
is not the same as 
that to find a particle inter-filament.
One more constitutive feature of a nematic phase is its fluidity.
One would expect that 
particles travel along the ``modulation'' 
in a ``modulated'' nematic phase 
as fast as they do 
along the nematic director 
in an ordinary nematic phase. 
This is what happens in the screw--like nematic phase that 
forms in systems of helical particles \cite{eliche}.
Preliminary results on the mechanism of diffusion in 
the filamentary phase that forms
in systems of hard infinitesimally--thin circular arcs
point that this is not the case \cite{diffilam}:
Hard infinitesimally--thin circular arcs  intrude from 
one into an adjacent filament in a step-like manner 
while retaining their orientation and rapidly return
to the original filament or advance to the subsequent filament:
A mechanism of diffusion that reminds the one operative
in a smectic A phase \cite{diffusme}.
These considerations are consistent with what has been ultimately observed in
systems of hard arched particles in three dimensions.
Initially, numerical simulations and experiments 
on systems of hard or colloidal arched particles 
have claimed that these systems form a splay-bend nematic phase \cite[(a,b)]{utrecht}.
Subsequently, this conclusion has been rectified:
That ``modulated'' phase is not positionally uniform:
it is not nematic but smectic-like \cite[(c)]{utrecht}. 
 

Rather,
the concavity and polarity of the present hard particles induces 
the recognition of another resemblance: 
between 
the cluster isotropic phase and the filamentary phase that 
are observed in section ({\sc{ii}}) of the present phase diagram 
and 
the ``living polymeric'' phase that 
was observed in systems of dipolar hard circles (discs)
\cite{dipolar}.
In both present and previous systems, 
filaments or chains form which are tortuous and 
occasionally ramificates and closes up to produce 
roundish clusters or irregular rings.

Sections ({\sc{i}}) and ({\sc{ii}}) of 
the phase diagram of systems of hard infinitesimally--thin 
circular arcs constitute (nothing else than)
the two--dimensional version of 
the phase diagram of systems of hard spherical caps 
with subtended angle $\uptheta \in [0,\uppi]$ \cite{calotta}.
In particular, 
the present filamentary phase is (nothing else than)
the two--dimensional version of that cluster columnar phase that 
was observed in three--dimensional systems of hard spherical caps.
Consistently to its lower dimension,
the filamentary phase is subject to 
stronger fluctuations that 
should conduce to 
a more extensive tortuosity 
as well as to 
more ramifications and ``ruptures''.
Yet, the same auto-assembly phenomenology is essentially observed 
in both two and three dimensions.
In particular, it causes 
the isotropic phase exhibit,
at sufficiently high $\uprho$,
the formation of clusters that
progressively pass 
from being filamentous or lacy 
to being roundish or globular
as  $\uptheta$ increases.
The principal difference between 
what is observed in two dimensions and 
what is observed in three dimensions is 
the capability of the two--dimensional roundish clusters 
to organise in a triangular lattice:
i.e., the formation of a cluster hexagonal phase in two dimensions. 

Though already incipient in section ({\sc{iii}}),
this cluster hexagonal phase completely characterises 
section ({\sc{iv}}) of 
the phase diagram of Fig. \ref{lafiguradiafase}. 
Consistently to the recent determination of 
the corresponding densest--known packings \cite{hcadkp},
it constitutes the truly novel phase that 
is observed in the present work.
It forms in two dimensions while
in three dimensions an analogous phase has not been observed and 
will probably be not observable. 
It forms due to the capability that 
hard infinitesimally--thin, particularly major, circular arcs have 
to intertwine without intersecting. 
This capability cannot be retained on going from two to three dimensions.

Similarly to what has been already commented
apropos the phase behavior of systems of hard spherical caps
\cite{calotta}, 
the auto-assembly phenomenology that is observed in 
systems of hard infinitesimally--thin circular arcs
resembles 
what is observed in molecular systems that form micelles \cite{micelle,Janus}:
These supramolecular structural units can be cylindrical or globular 
which can then auto-assemble to form a variety of complex phases
among which are columnar and crystalline phases \cite{micelle,Janus}.
There, at the origin of the complex phase behavior are
complex molecules that interact between them via
complicated attractive and repulsive intermolecular interactions.
Here, this complex phase behavior occurs in 
systems of relatively simple hard particles and 
thus is purely entropy--driven.

\section{conclusion and perspective}
\label{conclusion}
This work consists in an investigation on 
the phase behavior of 
systems of hard infinitesimally--thin circular arcs 
in the two--dimensional Euclidean space $\mathbb{R}^2$.
Depending on 
the subtended angle $\uptheta$
and 
the number density $\uprho$,
several purely entropy-driven phases are observed 
in the course of Monte Carlo numerical simulations 
\cite{MCorigin,MCWood,MClibri}.

Leaving aside the (quasi)nematic phase 
that is solely observed 
for sufficiently small values of $\uptheta$
and at intermediate values of $\uprho$,
more interesting are the other phases that are observed.
The very same isotropic phase is such:
Provided $\uprho$ is sufficiently high, 
it becomes no ordinary in that 
it exhibits clusters which pass 
from being filamentous to being roundish 
as $\uptheta$ progressively increases.
Provided $\uprho$ is even higher,
these two types of clusters respectively produce 
a filamentary phase
for $\uptheta \lesssim \uppi$
and  
a hexagonal phase for $\uptheta \gtrsim \uppi$. 
Both these phases are characterised by 
a ``supraparticular''  structural organisation:  
the actual structural units are formed by 
a number of suitably disposed hard particles. 
Particularly interesting is 
the cluster hexagonal phase.  
It offers examples of 
(soft) porous crystalloid materials \cite{vista,softporo} 
whose porosity could be regulated by compression.
The  auto--assembly phenomenology 
in systems of hard infinitesimally--thin circular arcs,
as well as that in 
systems of hard spherical caps \cite{calotta},
interestingly resemble that 
that occurs in micellising molecular systems 
\cite{micelle,Janus}.

Despite the extensive Monte Carlo numerical simulations,
there are several issues that 
still necessitate a clarification.
Leaving aside the persistent issue of
the nature of the nematic phase in 
a 
realistic 
two--dimensional system
\cite{straley},
it is the structure of the three cluster phases
that would require 
a more detailed characterisation.
This should centre on 
a detailed statistical analysis of 
the shape and size of respective clusters.
Specifically in 
systems of hard infinitesimally--thin minor circular arcs, 
it should address
the persistence length of the filaments and
the number of their ramifications and ``ruptures'' \cite{commentucolo}.
Specifically in 
systems of hard infinitesimally--thin major circular arcs, 
it should ascertain 
whether, on progressive compression, 
a single hexagonal phase forms 
or 
multiple hexagonal phases form 
in the (long) way towards 
the corresponding densest--known packings \cite{hcadkp}.
The clarification of these pending issues  necessitates to consider 
systems of a significantly larger size and 
improved computational resources and techniques
than those presently available.
Under these conditions, 
it would be very beneficial to have
an experimental system, either colloidal or granular, of 
thin-circular-arc--shaped particles 
\cite{mason}.
It could be used 
to first test  
the present predictions on the complete phase behavior
and 
then address and aid to resolve those pending issues. 

\section*{acknowledgments}
The authors acknowledge the support of
the Government of Spain 
under grant number FIS2017-86007-C3-1-P.


\begin{thebibliography}{99}
\bibitem{chaikin} 
e.g. P. M. Chaikin and T. C. Lubensky,
\textit{Principles of Condensed Matter Physics},
Cambridge University Press, Cambridge (1995).
\bibitem{torquato}
e.g. S. Torquato, 
\textit{J. Chem. Phys.} \textbf{149}, 020901 (2018).
\bibitem{crisplastici}
e.g. \textit{The Plastically Crystalline State:
Orientationally--Disordered Crystals},
edited by J. N. Sherwood, Wiley, Chichester (1979).
\bibitem{crisliquidi}
e.g. L. M. Blinov, 
\textit{Structure and Properties of Liquid Crystals},
Springer, Dordrecht (2011).
\bibitem{mvmrrev}
L. Mederos, E. Velasco, Y. Mart\'inez-Rat\'on,
\textit{J. Phys. Cond. Matter} \textbf{26}, 463101 (2014).
\bibitem{allenrev}
M. P. Allen, \textit{Mol. Phys.} \textbf{117}, 2391 (2019).
\bibitem{escobedo}
e.g. C. Avenda\~{n}o and F. A. Escobedo,
\textit{Curr. Op. Coll. Interface Sci.} \textbf{30}, 62 (2017). 
\bibitem{calotta}
(a) G. Cinacchi and J. S. van Duijneveldt, 
\textit{J. Phys. Chem. Lett.} \textbf{1}, 787 (2010);
(b) G. Cinacchi, 
\textit{J. Chem. Phys.} \textbf{139}, 124908 (2013);
(c) G. Cinacchi and  A. Tani, 
\textit{J. Chem. Phys.} \textbf{141}, 154901 (2014).
\bibitem{pine}
(a) S. Sacanna, M. Korpics, 
K. Rodriguez, L. Col\'on-Mel\'endez, 
S. -H. Kim, D. J. Pine, and G. -R. Yi,
\textit{Nature Comms.} \textbf{4}, 1688 (2013);
(b) K. V. Edmond, 
T. W. P. Jacobson, J. S. Oh,
G. -R. Yi, A. D. Hollingsworth,  S. Sacanna 
and
D. J. Pine,
\textit{Soft Matter} \textbf{17}, 6176 (2021). 

\bibitem{hcadkp} 
J. P. Ram\'irez Gonz\'alez and G. Cinacchi,
\textit{Phys. Rev. E} \textbf{102},  042903 (2020).
\bibitem{spiega} 
$\left [ x \right]$ indicates 
the strict floor function of the real variable $x$. 
The adjective strict is used to signify that 
this floor function returns $x-1$ and 
not $x$ for $x\in{\mathbb{N}}$ while 
it behaves as 
an ordinary floor function 
for $x \notin {\mathbb{N}}$.



\bibitem{MCorigin}
N. Metropolis, A. W. Rosenbluth, M. N. Rosenbluth, A. N. Teller, and E. Teller,
\textit{J. Chem. Phys} \textbf{21} , 1087 (1953).
\bibitem{MCWood}
W. W. Wood, \textit{J. Chem. Phys.} \textbf{48}, 415 (1968);
\textit{ibidem} \textbf{52}, 729 (1970).
\bibitem{MClibri}
e.g. 
(a)  M. P. Allen and D. J. Tildesley,
\textit{Computer Simulation of Liquids},
Clarendon Press, Oxford (1987);
(b) W. Krauth,
\textit{Statistical Mechanics:
Algorithms and Computations},
Oxford University Press, Oxford (2006).
\bibitem{O2}
Except for $\uptheta=2 \uppi$,
in which case it is $\mathsf{O}(2)$.

\bibitem{vieillard}
J. Vieillard-Baron, 
\textit{Mol. Phys.} \textbf{28}, 809 (1974).
This work actually introduced the method,
based on ordinary linear algebra,
to calculate 
$\mathtt{S}_2$ and ${\hat{\mathit{n}}}$ 
for 
a system of hard non-spherical particles 
in three dimensions. 
The method is,
mutatis mutandis, the same 
in a system of hard non--circular particles
in two dimensions.


\bibitem{comment1} 
Higher--order orientational pair--correlation functions and
their associated order parameters 
could be considered 
but,
based on the experience that was acquired with
systems of hard spherical caps 
\cite{calotta},
those of order one and two were considered sufficient.
In addition, more general orientational pair--correlation functions that depend 
not only on the modulus 
but also on the direction of 
the inter--particle distance vector
could be profitably considered. 

\bibitem{quasi}
The pre-fix (quasi) is added 
to indicate that, 
in a two--dimensional system, 
a proper long-ranged nematic ordering,
like any other proper long-ranged ordering,
would not exist.

\bibitem{comment2}
Orientational pair--correlation functions similar to 
Eqs. \ref{laG1orienta} and \ref{laG2orienta} 
could be defined in terms of 
the centres of the parent circles but 
their calculation was omitted.


\bibitem{discretegeometry}
e.g., P. Brass, W. O. J. Moser, J. Pach,
\textit{Research Problems in Discrete Geometry},
Springer, New York (2005).

\bibitem{mt19937}
M. Matsumoto and T. Nishimura, 
\textit{ACM Trans. Model. Comput. Simul.} 
\textbf{8}, 3 (1998).

\bibitem{blocks}
H. Flyvbjerg, H. G. Petersen,
\textit{J. Chem. Phys.} \textbf{91}, 461 (1989);
[consult also M. Jonsson,
\textit{Phys. Rev. E} \textbf{98}, 043304 (2018)].

\bibitem{onsager}
L. Onsager, 
\textit{Ann. N. Y. Acad. Scie.}  \textbf{51}, 627 (1949).

\bibitem{kayser}
R. F. Kayser  and H. J. Ravech\'{e},
\textit{Phys. Rev. A}  \textbf{17}, 2067 (1978).

\bibitem{berezinskii}
(a) V.L. Berezinskii,
\textit{Zh. Eksp. Teor. Fiz.} \textbf{59}, 907 (1970)
[\textit{Sov. JETP} \textbf{32}, 493 (1971)].
(b) V.L. Berezinskii,
\textit{Zh. Eksp. Teor. Fiz.} \textbf{61}, 1144 (1971)
[\textit{Sov.JETP} \textbf{34}, 610 (1972)].
(c) V.L. Berezinskii, 
''Nizkotemperaturnye svoistva dvumernykh sistem s nepreryvnoi gruppoi simmetrii'',
Ph.D. thesis, Landau Institute for Theoretical Physics, Moscow, 1971;
\textit{Nizkotemperaturnye svoistva dvumernykh sistem s 
nepreryvnoi gruppoi simmetrii},
Fizmalit, Moscow (2007).

\bibitem{kt}
(a) J. M. Kosterlitz and D. J. Thouless,
\textit{J. Phys. C} \textbf{5}, L124 (1972);
(b) J. M. Kosterlitz and D. J. Thouless,
\textit{J. Phys. C} \textbf{6}, 1181 (1973);
(c) J. M. Kosterlitz, 
\textit{J. Phys. C} \textbf{7}, 1046 (1974).

\bibitem{ktrev}
J. M. Kosterlitz, 
\textit{Rep. Prog. Phys.} \textbf{79}, 026001 (2016).

\bibitem{ktrus}
V. N. Ryzhov, E. E. Tareyeva, Y. D. Fomin, E. N. Tsiok,
\textit{Usp. Fiz. Nauk.} \textbf{187}, 921 (2017) 
[\textit{Phys. Usp.} \textbf{60}, 857 (2017)].

\bibitem{eppenga}
D. Frenkel and R. Eppenga
\textit{Phys. Rev. A} \textbf{31}, 1776 (1985).

\bibitem{india}
M. D. Khandkar and M. Barma,
\textit{Phys. Rev. E} \textbf{72},  051717 (2005).

\bibitem{vink}
R. L. C. Vink
\textit{Eur. Phys. J. B} \textbf{72},  225 (2009).

\bibitem{polonia}
A. Chrzanowska, \textit{Acta Phys. Pol. B} \textbf{36}, 3163 (2005). 

\bibitem{mermin1}
N. D. Mermin and H. Wagner,
\textit{Phys. Rev. Lett.} \textbf{17}, 1133 (1967).

\bibitem{mermin2}
N. D. Mermin,
\textit{Phys. Rev.} \textbf{176}, 250 (1968).

\bibitem{straley}
J. P. Straley, \textit{Phys. Rev. A} \textbf{4}, 675 (1971).

\bibitem{fvt1}
J. G. Kirkwood, 
\textit{J. Chem. Phys.} \textbf{18}, 380 (1950);
W. W. Wood, 
\textit{J. Chem. Phys.} \textbf{20}, 1334 (1952);
Z. W. Salsburg and W. W. Wood,
\textit{J. Chem. Phys.} \textbf{37}, 798 (1962);
F. H. Stillinger and Z. W. Salsburg,
\textit{J. Stat. Phys.} \textbf{1},  179 (1969).

\bibitem{fvt2}
A. Donev, S. Torquato and F. H. Stillinger,
\textit{Phys. Rev. E} \textbf{71}, 011105 (2005).


\bibitem{planarot1985}
D. H. Lee and G. Grinstein,
\textit{Phys. Rev. Lett.} \textbf{55}, 541 (1985).

\bibitem{planarot1989}
D. B. Carpenter and J. T. Chalker,
\textit{J. Phys. Condens. Matter} \textbf{1}, 4907 (1989).

\bibitem{planarot2011}
(a) F. C. Poderoso, J. J. Arenzon, Y. Levin,
\textit{Phys. Rev. Lett.} \textbf{106}, 067202 (2011);
(b) G. A. Canova, Y. Levin, J. J. Arenzon,
\textit{Phys. Rev. E} \textbf{89}, 012126 (2014).
(c) G. A. Canova, Y. Levin, J. J. Arenzon,
\textit{Phys. Rev. E} \textbf{94}, 032140 (2016).

\bibitem{planarot2012} 
D. M. H\"ubscher and S. Wessel,
\textit{Phys. Rev. E} \textbf{87}, 062112 (2012). 

\bibitem{zukovic}
In this respect, consult:
(a) M. \v{Z}ukovi\v{c}, G. Kalagov,
\textit{Phys. Rev. E} \textbf{96}, 022158 (2017);
(b) M. \v{Z}ukovi\v{c}, G. Kalagov, 
\textit{Phys. Rev. E} \textbf{97}, 052101 (2018);
(c) M. \v{Z}ukovi\v{c},
\textit{Phys. Lett. A} \textbf{382}, 2618 (2018).

\bibitem{pv}
H. Chamati and S. Romano,
\textit{Phys. Rev. E} \textbf{77}, 051704 (2008). 

\bibitem{zigzag}
(a) R. Tavarone, P. Charbonneau, H. Stark,
\textit{J. Chem. Phys.} \textbf{143}, 114505 (2015);
(b) P. Karbowniczek, 
\textit{J. Chem. Phys.} \textbf{148}, 136101 (2018).

\bibitem{eliche}
(a) E. Barry, Z. Hensel, Z. Dogic, M. Shribak and R. Oldenbourg,
\textit{Phys. Rev. Lett.} \textbf{96}, 018305 (2006); 
(b) H.B. Kolli, E. Frezza, G. Cinacchi, A. Ferrarini, A. Giacometti,  T.S. Hudson,
\textit{J. Chem. Phys.} \textbf{140}, 081101 (2014);
(c) G. Cinacchi, A.M. Pintus, A. Tani,
\textit{J. Chem. Phys.} \textbf{145}, 134903 (2016).

\bibitem{diffilam}
J.P. Ram\'irez Gonz\'alez, G. Cinacchi, in progress.

\bibitem{diffusme}
(a) M.P. Lettinga, E. Grelet,
\textit{Phys. Rev. Lett.} \textbf{99}, 197802 (2007);
(b) G. Cinacchi, L. De Gaetani,
\textit{Phys. Rev. E} \textbf{79}, 011706 (2009).

\bibitem{utrecht}
(a) M. Chiappini, T. Drwenski, R. van Roij, M. Dijkstra,
\textit{Phys. Rev. Lett.} \textbf{123}, 068001 (2019);
(b) C. Fern\'andez-Rico, M. Chiappini, T. Yanagishima, H. de Sousa, D.G.A.L. Aarts, M. Dijkstra, R.P.A. Dullens,
\textit{Science} \textbf{369}, 950 (2020);
(c) M. Chiappini, M. Dijkstra, 
\textit{Nature Communications} \textbf{12}, 2157 (2021).

\bibitem{dipolar}
J.M. Caillol, J.J. Weis,
\textit{Mol. Phys.} \textbf{113}, 2487 (2015).  

\bibitem{micelle}
e.g. J.N. Israelachvili,  
\textit{Intermolecular and Surface Forces},
Academic Press (2011).

\bibitem{Janus}
Other systems for which a certain resemblance could be recognised
are colloidal systems that are formed by Janus colloidal particles:
e.g. J.S. Oh, S. Lee, S.C. Glotzer, G.-R. Yi, D.J. Pine,
\textit{Nature Comms.} \textbf{10}, 3936 (2019).

\bibitem{vista}
If such a view is naturally taken along 
the direction that is perpendicular to $\mathbb{R}^2$.

\bibitem{softporo}
e.g. A.G. Slater and A.I. Cooper,
\textit{Science} \textbf{348}, aaa8075 (2015).

\bibitem{commentucolo}
it is not clear whether these ramifications and ``ruptures''should be considered as ``defects'' or 
are instead inherent to the very nature of these phases.

\bibitem{mason}
One possibility could be systems of colloidal particles that
suitably generalised 
those that have been prepared and investigated by
P. Y. Wang and T. G. Mason, \textit{J. Am. Chem. Soc.} \textbf{137},
15308 (2015).

\end{thebibliography}
\end{document}